\documentclass[num-refs]{wiley-article}

\usepackage{subfigure}
\usepackage{siunitx}
\usepackage{gensymb}
\usepackage[final]{pdfpages}

\papertype{Original Article}
\paperfield{Journal Section}

\title{Reduced etch lag and high aspect ratios by deep reactive ion etching (DRIE)}

\abbrevs{ICP, inductively coupled plasma, DRIE, deep reactive ion etching, PR, photoresist.}

\author[1\authfn{1}]{Michael S. Gerlt}
\author[2\authfn{1}]{Nino F. Läubli}
\author[1]{Michel Manser}
\author[2]{Bradley J. Nelson}
\author[1]{Jürg Dual}

\contrib[\authfn{1}]{Equally contributing authors.}

\affil[1]{Department of Mechanical and Process Engineering, Institute for Mechanical Systems, Zurich, 8092, Switzerland}
\affil[2]{Department of Mechanical and Process Engineering, Institute of Robotics and Intelligent Systems, Zurich, 8092, Switzerland}

\corraddress{Michael Gerlt, Department of Mechanical and Process Engineering, Institute for Mechanical Systems, Zurich, 8092, Switzerland}
\corremail{gerlt@imes.mavt.ethz.ch}

\fundinginfo{This work is supported by the ETH Zurich and, in part, an interdisciplinary grant from the Swiss National Science Foundation (Grant Number CR22I2\_166110).}

\runningauthor{Michael S. Gerlt et al.}

\begin{document}

\begin{frontmatter}
\maketitle

\begin{abstract}
Deep reactive ion etching (DRIE) with the Bosch process is one of the key procedures used to manufacture micron-sized structures for MEMS and microfluidic applications in silicon and, hence, of increasing importance for miniaturization in biomedical research. While guaranteeing high aspect ratio structures and providing high design flexibility, the etching procedure suffers from reactive ion etching lag and often relies on complex oxide masks to enable deep etching. In this work, we introduce an optimized Bosch process that reduces the etch lag to below $\SI{1.5}{\percent}$. Furthermore, we improved a three-step Bosch process, allowing the fabrication of structures with $\SI{6}{\micro\meter}$ thickness at depths up to $\SI{180}{\micro\meter}$ while maintaining their stability.

\end{abstract}
\end{frontmatter}

\section{Introduction}\label{sec:int}
\begin{figure}[t!]
\centering
    \includegraphics[width=\linewidth/2]{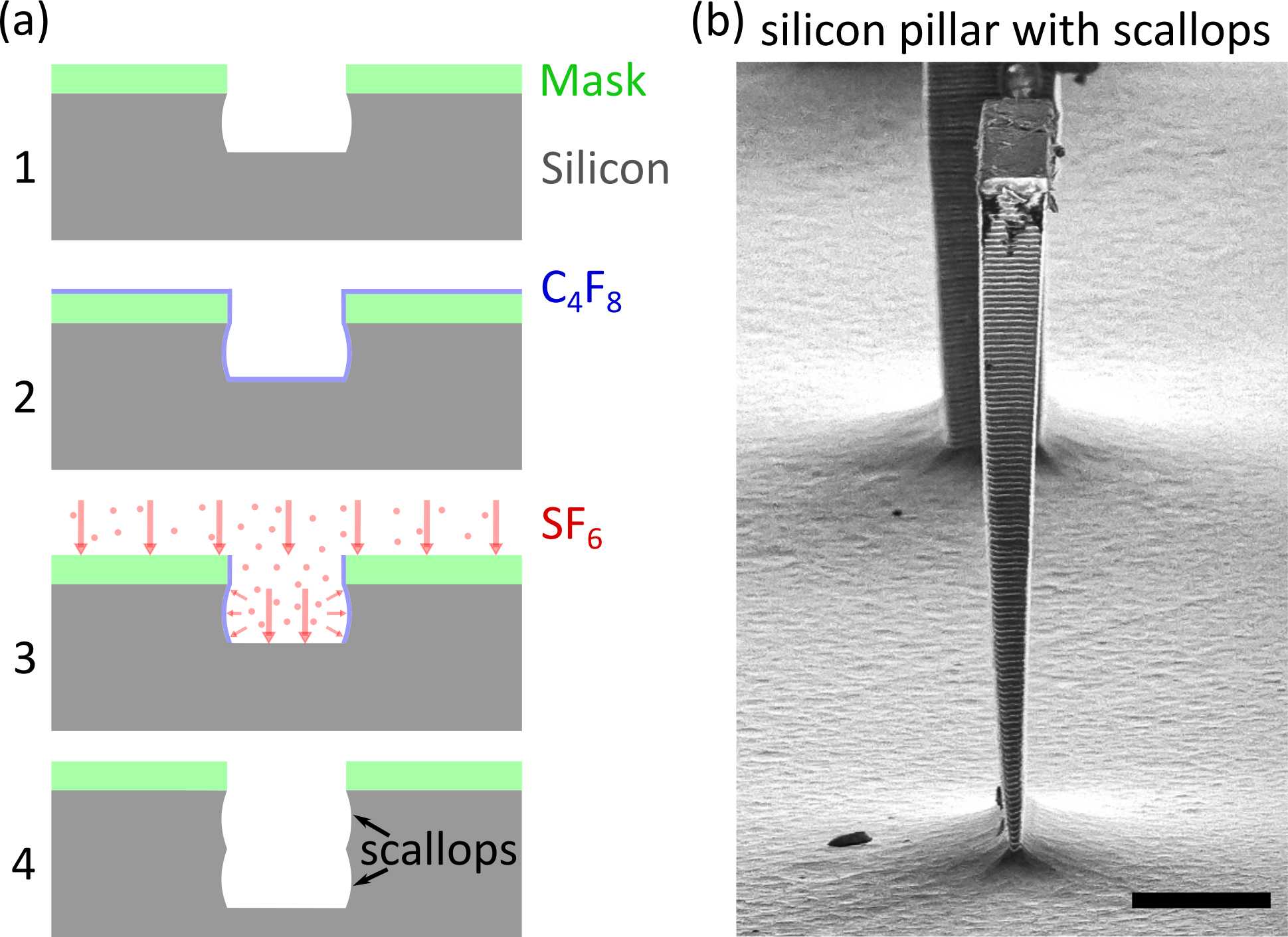}
    \caption{\textbf{Bosch process.} 
    (a) Sketch of the Bosch process. (1) Silicon wafer with mask after one process cycle. (2) Deposition of a polymer layer during passivation. (3) Physical and chemical etching. (4) Silicon wafer with mask after two process cycles. (b) Scanning electron microscopy image of a $\SI{10}{}$x$\SI{40}{\micro\meter}$ silicon pillar. The scallops that result from the sidewall etching are clearly visible. Scale bar corresponding to $\SI{20}{\micro\meter}$.
    }
    \label{fgr:Fig0}
\end{figure}

With the rise of the semiconductor industry, cleanroom-based production technologies in silicon gained a significant increase in popularity. The ability to fabricate micron-sized structures has revolutionized numerous industries and led to previously unseen products such as MEMS\cite{10.1109/JPROC.2003.820534} and microfluidic devices\cite{10.1038/nature05058} that recently attained increasing relevance for biotechnology and biomedical research.\cite{10.1021/ac0605602} 
The standard fabrication steps used for the production of these microstructures consist of photolithography, etching, and post-processing such as bonding, dicing, and packaging. Given the high complexity of the etching step in current microfabrication processes, various approaches have been developed, each having their advantages and disadvantages. One of the most common processes for microstructure fabrication is deep reactive ion etching (DRIE), where physical and chemical etching are successfully combined.\cite{10.1063/1.3474652} The full potential of DRIE was revealed with the invention of a time-multiplexed alternating process of passivation and etching by Laemer and Schilp in 1996.\cite{BoschPatent} Named after their employer, the Bosch process is nowadays one of the key processes in the silicon industry.\cite{10.1002/ppap.201800207} During the passivation step, a thin polymer layer, typically consisting of octafluorocyclobutan ($C_4F_8$), is deposited onto the substrate, while in the subsequent etching step, the passivation layer and the underlying silicon are etched. Due to the ions' acceleration towards the target, significantly higher vertical etch rates can be achieved in sidewall etching, leading to the formation of high aspect ratio trenches (Fig. \ref{fgr:Fig0} (a)). The characteristic scallops that result from the Bosch process are unavoidable. However, given their size in the nanometre-range, the influence of the scallops is usually negligible (Fig. \ref{fgr:Fig0} (b)).

One of the Bosch process's fundamental limitations is the reactive ion etching (RIE) lag, which leads to narrow features being etched shallower than large features (RIE lag) as the gas dynamics limit the gas transport to and from structures with an aspect ratio exceeding 1:1.\cite{10.1016/B978-0-12-817786-0.00016-5} Recently, several research groups tried to tackle the RIE lag by adding process steps or advanced timing controls, which often significantly increases process complexity and, thus, requires expensive tools.\cite{10.1007/s00542-006-0211-2, 10.1088/0960-1317/14/4/029, 10.1116/1.2172944, 10.1016/S0169-4332(01)00197-0} In this work, we achieved a RIE lag reduction to below 1.5\% at an etch depth of $\SI{50}{\micro\meter}$ by solely adjusting parameters of a two-step Bosch process while maintaining its simplicity and adaptability for a large variety of fabrication machines.

Another limitation of the two-step Bosch process is its low selectivity, which complicates the production of structures with high aspect ratios.\cite{10.1016/S0167-9317(03)00089-3} Therefore, several approaches to optimize the selectivity, \textit{i.e.}, the ratio between mask and silicon etch, were developed.\cite{10.1016/j.sna.2007.12.026, 10.1016/j.mee.2013.06.010, 10.1088/0957-4484/19/34/345301} Nevertheless, for a reliable etch depths of more than $\SI{100}{\micro\meter}$, silicon oxide masks are required, which demands an additional patterning step in a Reactive Ion Etching (RIE) machine.\cite{10.1109/MEMSYS.2012.6170138, 10.1116/1.1415511} 
To overcome this challenge, we introduce a three-step Bosch process, in which the duration of the second process step (anisotropic etching) is lowered to the time necessary to remove the polymer layer on the bottom of the trench, \textit{i.e.}, breakthrough step. Additionally, a third step is introduced, which does not rely on particle acceleration towards the target, thus reduces the etch rate of the mask and increases selectivity.
Our improvement of the three-step Bosch process enabled us to use a photoresist with $\SI{1.4}{\micro\meter}$ thickness to etch depths greater than $\SI{450}{\micro\meter}$ corresponding to a selectivity of more than 350. Furthermore, our results illustrate etching angles near $\SI{89.7}{\degree}$, which allows for the production of a silicon wall with $\SI{6}{\micro\meter}$ thickness in between two microchannels to a depth of more than $\SI{270}{\micro\meter}$ while maintaining its impermeability, crucial for subsequent use in microfluidics.

In this work, we present an optimized two and three-step Bosch process that tackles two important challenges of DRIE etching, i.e. the RIE lag and the selectivity. We believe that our achievements concerning the well-described and characterized processes simplify the adaptability and accessibility of state-of-the-art technologies for a broader audience.
\section{Results \& Discussion}\label{sec:RD}
\subsection{Bosch process optimization for reduced RIE lag}

\begin{figure}[b!]
\centering
    \includegraphics[width=\linewidth]{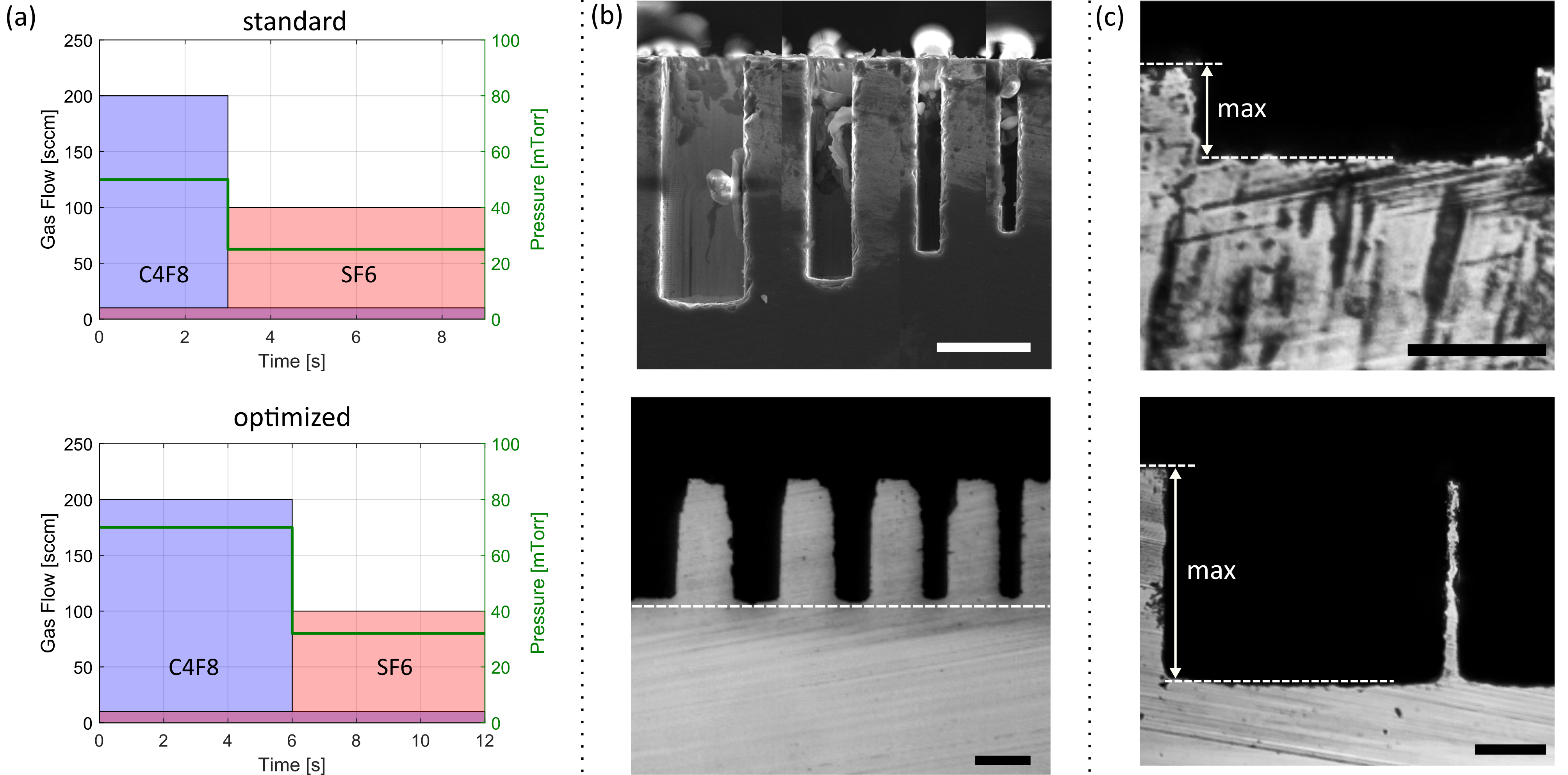}
    \caption{\textbf{Reduced RIE lag.}
    (a) Sketch of the Bosch process parameters for a standard procedure (top) and the optimized process parameters (bottom). A gas flow of 10 sscm is always maintained to enable faster switching times (violet horizontal bar). (b) Optical microscopy images of diced silicon wafers to illustrate the improvement in RIE lag. (top) Trenches etched with the standard Bosch process, with $\SI{20}{\micro\meter}$ to $\SI{3}{\micro\meter}$ wide trenches reached a depths of $\SI{64}{\micro\meter}$ and $\SI{47}{\micro\meter}$, respectively, corresponding to an RIE lag of $\SI{27}{\percent}$. (bottom) Trenches etched with our optimized process, with $\SI{20}{\micro\meter}$ to $\SI{5}{\micro\meter}$ width were etched $\SI{47.8}{} \pm \SI{1.9}{\micro\meter}$ deep into the silicon wafer, corresponding to a RIE lag of below $\SI{1.5}{\percent}$. Scale bars corresponding to $\SI{25}{\micro\meter}$.
    (c) Microscope pictures of a diced silicon wafer with a $\SI{200}{\micro\meter}$ wide trench. The maximal achievable etch depth with a $\SI{1.4}{\micro\meter}$ PR layer (top) and a $\SI{2}{\micro\meter}$ $SiO_2$ mask (bottom) as the mask was $\SI{29}{\micro\meter}$ and $\SI{141}{\micro\meter}$, respectively. Scale bar corresponding to $\SI{50}{\micro\meter}$.
    }
    \label{fgr:Fig1}
\end{figure}

The Bosch process is one of the most common DRIE processes as it consists of only two steps while not relying on fast response times only attainable with high-end equipment, making it accessible for a broader audience and a large variety of research fields. However, despite its simplistic approach, the process is capable of providing etch angles close to $\SI{90}{\degree}$ or even higher, enabling the production of small structures while guaranteeing their stability on the channel bottom, \textit{i.e.}, inside the trenches.
Nevertheless, one major limitation of the Bosch process is the RIE lag where channels with different widths are not etched to the same depth due to the limited gas supply within the constricted features, which can already be relevant for channels with widths smaller than $\SI{100}{\micro\meter}$.

We successfully overcame this limitation following a concept initially introduced by Lai \textit{et al.}, focusing on adjusting the ratio between the duration of the passivation process $t_P$ and etching process $t_E$.\cite{10.1116/1.2172944} While the Bosch process only consists of two steps, one process cycle can be separated into three individual tasks, \textit{i.e.}, polymer deposition, polymer etching, and silicon etching (see Supporting Figure S1). It is essential to highlight that the polymer deposition and the silicon etching are primarily chemical and, by that, diffusion dependent, while the initial polymer etching at the bottom of the trench is of physical nature and, therefore, mainly independent of the aspect ratio.

To better understand why a substantial increase of the passivation time, in comparison to a standard Bosch process (top of Fig \ref{fgr:Fig1} (a)), led to a significant reduction in RIE lag, we have a brief look at the procedure for a wide as well as a narrow trench. In narrow structures, the polymer deposition is slower than in wide features, leading, for a given deposition duration, to a thinner passivation layer at the confined space compared to open spaces. The subsequent anisotropic physical polymer etching, however, is independent of the aspect ratio. Therefore, the silicon in narrow trenches will be exposed faster than for a wide area due to the thinner passivation layer. The last step, \textit{i.e.} the chemical silicon etching, is again slower in confined spaces. However, as it starts earlier than in wide structures, due to the silicon at the bottom being exposed faster, the process duration can be chosen such that, after a single cycle of the Bosch process, the wide and the narrow trenches end at the same depth. Following the assimilation of the two process durations $t_P$ and $t_E$, slight adjustments have been performed through the iterative increment of the chamber pressure during the etching step to improve reliability and uniformity further. Please refer to Tables 1 and 2 in the Supporting Information for a detailed overview of all processes' parameters.

With these adjustments to the original process (bottom of Fig. \ref{fgr:Fig1} (a)), we were able to etch channels with a width ranging from $\SI{5}{\micro\meter}$ to $\SI{20}{\micro\meter}$ down to a depth of $\SI{48}{\micro\meter}$ while the smallest channel showed a difference in height of less than $1.5 \%$ (Fig. \ref{fgr:Fig1} (b)). Despite the etch angle $\alpha = 91 \pm 0.5 \si{\degree} $ of our modified process, the etch depth is limited. With a PR layer of $\SI{1.4}{\micro\meter}$ thickness, we achieved a maximal etch depth of $\SI{29}{\micro\meter}$ corresponding to a selectivity of 22. When using a $SiO_2$ mask with $\SI{2}{\micro\meter}$ thickness, the maximal etch depth increased to $\SI{141}{\micro\meter}$ corresponding to a selectivity of 71 (Fig \ref{fgr:Fig1} (c)). Measurements of multiple structures from different wafers revealed reproducibility of our process with etch angles of $\SI{91.14}{\degree} \pm \SI{0.69}{\degree}$ (see Supporting Figure S4).
In summary, by adjusting the etch to deposition time ratio and reducing the pressure, we reduced the RIE lag to below $\SI{1.5}{\percent}$ and produced structures as deep as $\SI{141}{\micro\meter}$. It is important to highlight that our approach allowed for the controlled etching of wafers with large percentages of exposed area (as high as $\SI{90}{\percent}$), a task which typically provides an additional challenge in MEMS fabrication.

\subsection{High rate process optimization for deep etch}

\begin{figure}[t!]
\centering
    \includegraphics[width=\linewidth]{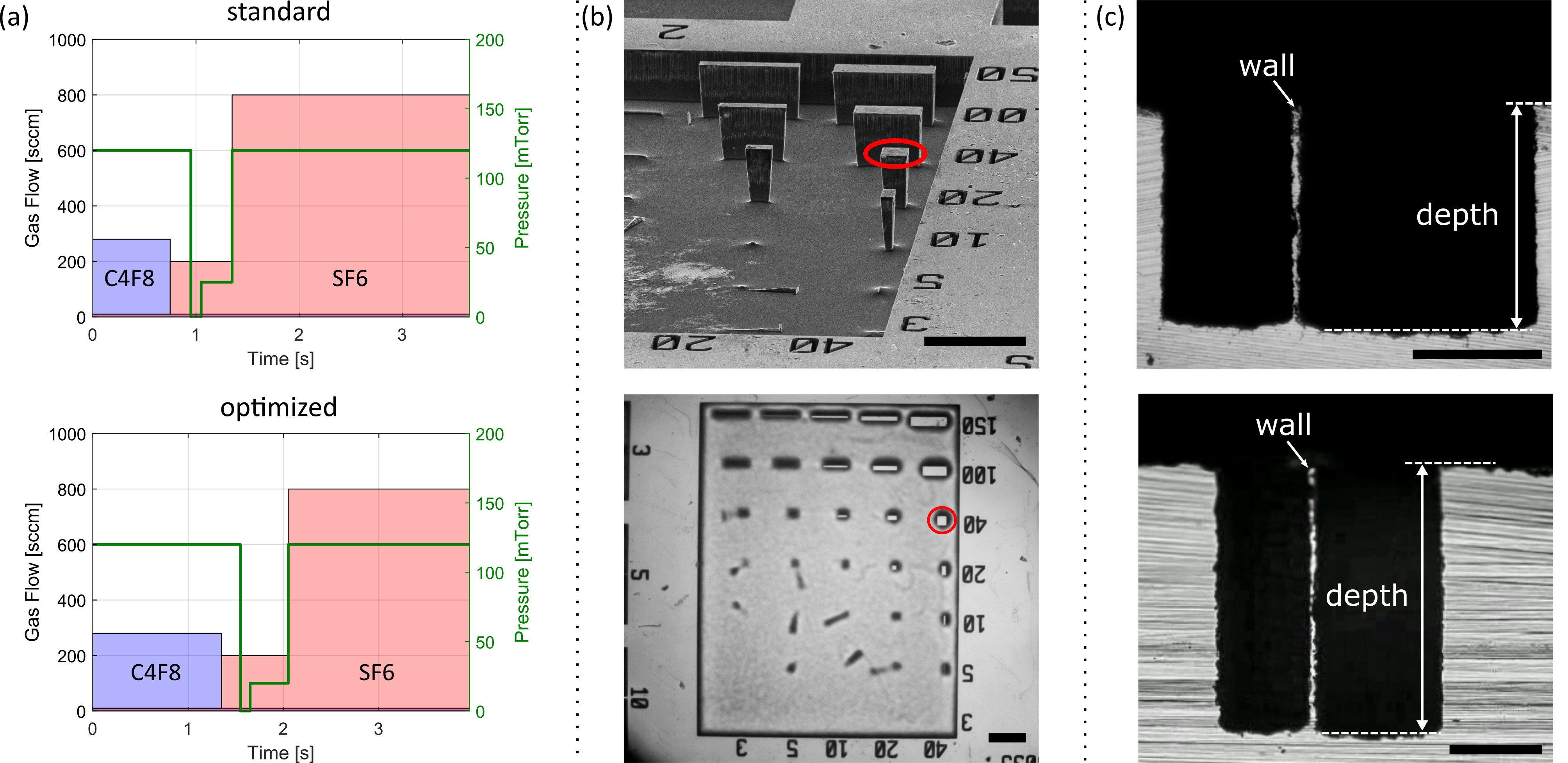}
    \caption{
    \textbf{High aspect ratio etching of small structures.}
    (a) Sketch of the standard three-step Bosch process parameters as provided by the supplier (top) and the optimized parameters (bottom).
    (b) Scanning electron microscopy (top) and optical microscopy (bottom) images of micron-sized pillars with $\SI{100}{\micro\metre}$ height, produced with the standard (top) and optimized (bottom) process parameters. Numbers on the side and the bottom correspond to the length and width of the structures, respectively. The same pillar has been marked with a red circle for demonstration purposes. Scale bars corresponding to $\SI{150}{\micro\meter}$.
    (c) Optical microscopy pictures of a diced silicon wafer with two channels and a thin wall in-between. We etched the tranches $\SI{186}{\micro\metre}$ (top) and $\SI{277}{\micro\metre}$ (bottom) deep into the silicon wafer. The thin wall is $6.4 \pm 0.2 \si{\micro\meter}$ wide (top) and $8.4 \pm 0.5 \si{\micro\meter}$ wide (bottom). The channels have an etch angle of $89.83 \degree$. Scale bar corresponding to $\SI{100}{\micro\meter}$.
    }
    \label{fgr:Fig2}
\end{figure}

Due to the limitations in the etch depth of the two-step Bosch process, we analyzed and optimized a three-step Bosch process for small high-aspect-ratio structures. Small structures typically rely on thin photoresist layers to ensure appropriate resolution during the preceding photolithography step as vertical resist walls are yet unobtainable, limiting the etch depth to around $\SI{100}{\micro\metre}$.
Therefore, the combination of deep etching with thin PR layers demands a process with high selectivity, such as achievable through the means of a three-step Bosch process. Here, the anisotropic etching step, where ions are accelerated towards the target with platen power, solely focuses on etching away the thin polymer layer formed on the bottom of the trench in the previous passivation step (Fig. \ref{fgr:Fig0} (a)). Therefore, the anisotropic etching step has a much shorter duration in comparison to the two-step Bosch process. A second etching step is introduced, in which the platen power, which is responsible for the ion and radical acceleration, is switched off, and the $SF_6$ gas-flow is increased by a factor of four. In this second etching step, the etching happens predominately chemical, leading to a significant increase in selectivity.
One of the main challenges of the three-step Bosch process is the fabrication of structures with etch angles below $\SI{90}{\degree}$, which sets a limit to the minimal producible feature size due to possible structural instabilities at the bottom of the etched features. The etch angle of a standard three-step Bosch process (Fig \ref{fgr:Fig2} (a)) is around $\SI{87}{\degree}$ - $\SI{88}{\degree}$, depending on the etch depth. 
Through precise adjustment of the anisotropic etching step and the chamber pressure, we achieved an etch angle of $\SI{89.6}{\degree} \pm \SI{0.13}{\degree}$ reproducible for various trench geometries (see Supporting Figure S5). To be more specific, we lowered the pressure during the second step, \textit{i.e.},  the breakthrough step, leading to a higher mean free path for the ions being accelerated towards the sample and thus decreasing the sidewall etch. Furthermore, we adjusted the passivation and etching times to our specific design, leading to an overall thicker polymer layer while still ensuring successful removal at the microchannels' bottom. Depending on the etched design, the passivation time might require individual adjustments to prevent black silicon formation. Please refer to Tables 3 and 4 in the Supporting Information for a detailed overview of all the processes' parameters.
It is important to note that all parameters shown in Figure \ref{fgr:Fig2} (a) rely on a short response time of the machine components and, therefore, high-end equipment might be necessary.
With the previously introduced adjustments, we were able to achieve an etch angle of $\SI{89.6}{\degree} \pm \SI{0.13}{\degree}$ enabling us to produce micrometer-sized pillars that maintained their stability even at $\SI{100}{\micro\meter}$ etch depth. Compared to a standard three-step Bosch process, where the smallest stable structure was a $\SI{40}{}$x$\SI{20}{\micro\meter}$ pillar (top of Fig. \ref{fgr:Fig2} (b)), we were able to produce a $\SI{40}{}$x$\SI{5}{\micro\meter}$ pillar with our improved process (bottom of Fig. \ref{fgr:Fig2} (b)), lowering the critical feature size by a factor of 4. 
Furthermore, our approach demonstrates suitability for microfluidic applications. For this purpose, we etched two microfluidic channels $\SI{186}{\micro\metre}$ deep into the silicon with a distance of only $\SI{6}{\micro\meter}$ resulting in a thin silicon wall (top of Figure \ref{fgr:Fig2} (c) ). To ensure the stability of the produced structures, we bonded glass to the top of the silicon wafer and applied a water flow of $\SI{7.5}{\milli\metre\per\second}$ through one of the microfluidic channels to prove the wall's impermeability (see Supporting Figure S2). By increasing the distance of the two channels to $\SI{8}{\micro\meter}$, we were able to increase the etch depth to $\SI{277}{\micro\metre}$ (bottom of Figure \ref{fgr:Fig2} (c) ).
Finally, we were able to etch more than $\SI{450}{\micro\meter}$ deep into the silicon with a thin PR layer of $\SI{1.4}{\micro\meter}$ (see Supporting Figure S3), illustrating the high flexibility of our approach as required for complex miniaturization in novel biomedical lab-on-a-chip devices.
\section{Conclusion}\label{sec:C}
In this work, we tackled two fundamental challenges of DRIE etching, namely the RIE lag and high aspect ratio etching of small structures, and described the optimized processes in detail. Through an improved two-step Bosch process, we were able to etch trenches of various widths within a difference in etch depth of less than $\SI{1.5}{\percent}$ while keeping the simplicity of the Bosch process to ensure the broad and well-established applicability of this procedure.
Further, we optimized a three-step Bosch process while focusing on the reproducibility of etch angles close to $\SI{90}{\degree}$ and improving the overall selectivity. With our refined process, we were able to etch a $\SI{40}{}x\SI{5}{}x\SI{100}{\micro\meter}$ pillar, lowering the critical feature size compared to the standard process by a factor of 4. Additionally, we etched a $\SI{6}{\micro\metre}$ thin wall in between two microchannels down to a depth of $\SI{186}{\micro\metre}$, while maintaining impermeability to water which is of substantial importance for subsequent lab-on-chip applications. 
Our novel and improved procedures are of interest to a vast number of research groups as they allow for the fabrication of novel structures and devices as well as improving the performance of already existing applications.
\section{Experimental Section}\label{sec:E}
All processes were carried out on standard single side polished silicon wafers ($500 \pm \SI{25}{\micro\meter}$ thickness, Prolog Semicor, Ukraine). To determine the limits in terms of etch depth for the Bosch-process, we thermally evaporated $\SI{2}{\micro\meter}$ $SiO_2$ on the polished side of some silicon wafers.

First, we processed the wafers in a yellow room with standard photolithography processes. To enhance the adhesion of photoresist (PR), we deposited hexamethyldisilazane (HMDS) on the wafers for $\SI{30}{\second}$. Then, $\SI{4}{\milli\litre}$ of PR was dropped on the polished side of the silicon wafers. We used two different resists, namely S1813 (Shipley, United Kingdom) and AZ nLOF 2070 (Microchemicals, Germany). We spin coated the resists at $\si{4000}$ r.p.m. for $\SI{35}{\second}$ onto the wafers resulting in a resist thickness of $1.4 \pm 0.1 \si{\micro\meter}$ and $7.2 \pm 0.2 \si{\micro\meter}$ for the S1813 and the AZ nLOF 2070, respectively. Next, the samples were soft-backed for $\SI{1}{\minute}$ per micrometre resist thickness at $\SI{100}{\degreeCelsius}$ to evaporate solvents in the PR layer. Afterwards, the wafers were re-hydrated within the cleanroom (40-50 \% humidity) for 10 minutes per micrometre.

After finishing the PR deposition, the samples were transferred to the Mask Aligner (MABA6, Süss Microtec, Germany) and exposed to UV-light from a mercury lamp. The exposure time depends on the exposure dose of the resists and the measured intensity of the lamp. S1813 at $\SI{1.4}{\micro\meter}$ thickness required an exposure dose of $\SI{230}{\milli\joule\per\square\centi\meter}$ and thus, given the UV lamps intensity of $\SI{7.4}{\milli\watt\per\square\centi\metre}$, needed to be exposed for $\SI{30}{\second}$ in our machine.  AZ nLOF 2070 at $\SI{7}{\micro\meter}$ thickness required an exposure dose of $\SI{170}{\milli\joule\per\square\centi\meter}$ and thus, needed to be exposed for $\SI{23}{\second}$ in our machine. Further, we set the contacting settings to hard vacuum.

Finally, we developed the wafers. AZ351B (Microchemicals, Germany) was diluted with a ratio of 1:5 developer to distilled water and has been utilized for the development of S1813. AZ 726 MIF (Microchemicals, Germany) has been used undiluted for the development of AZ nLOF 2070. The development was assisted by putting the Petri dish with developer solution and wafer into an ultrasonic bath (ultrasonic power set to level 3 of 10). The wafers were removed from the developer and put into a water bath to stop the development when the silicon or the silicon dioxide gets visible. This typically happened after $\SI{15}{\second}$ for S1813 and after $\SI{60}{\second}$ for AZ nLOF 2070. 

For the wafers covered with $SiO_2$, we transferred the samples into a Reactive Ion Etching (RIE) machine (PlasmaPro NGP80, Oxford Instruments, United Kingdom) and etched the complete $\SI{2}{\micro\meter}$ $SiO_2$ layer of the exposed areas. The process we used consisted of two steps that were repeated twelve times. The first step (etch) consisted of 5 minutes $CHF_3$ flow at $40$ sscm, $CF_4$ flow at $40$ sscm, $O_2$ flow at $5$ sscm and $\SI{130}{\watt}$ high frequency (HF, $\SI{13.56}{\mega\hertz}$) power. Then, all flows and the HF power were turned of for 5 minutes to let the sample cool down. All process steps were carried out at $\SI{15}{\degreeCelsius}$ and at 15 mTorr.

After photolithography or RIE etching, respectively, the samples were transferred into the deep reactive ion etching (DRIE) machine (PlasmaPro Estrelas100, Oxford Instruments, United Kingdom). The processes carried out here are well described in the result section of this publication.
Finally, the wafers were diced into small pieces with a dicing SAW (DAD$3221$, Disco corporation, Germany). We used a hub-blade (FTB R46 45130, Disco corporation, Germany) and diced at 30000 rpm with $\SI{0.5}{\milli\meter\per\second}$ to avoid damaging the small structures. However, as can be seen in Figure \ref{fgr:Fig1} (c), damage of the wafers' surface at the top of microfluidic channels could not be avoided completely.

For evaluation of the results, the diced samples were inspected under a microscope (Axioscope, Zeiss, Germany). We utilized a blue LED (505 nm) from the top and no backlight to achieve the highest possible contrast. We connected a camera with a one-inch sensor to the microscope and evaluated the sample at two different magnifications, 20x and 40x respectively.
Thin structures such as pillars tend to rip off during dicing due to the high momentum. These structures were inspected using a scanning electron microscope (Nova NanoSEM 450, Thermo Fisher Scientific, USA). We tilted the samples to examine the structures in all three spatial dimensions.

\section*{acknowledgements}
The authors would like to express their gratitude to Rasmus Pedersen from Oxford Instruments for his continous support regarding etch process optimisation. This work is supported by the ETH Zurich and, in part, by an interdisciplinary grant from the Swiss National Science Foundation (Grant Number CR22I2\textunderscore 166110) to B.J.N..

\section*{conflict of interest}
The authors declare no conﬂict of interest

\section*{supporting information}
Supporting Information is available from the author


\newpage
\bibliography{02_references}

\begin{thebibliography}{17}
\providecommand{\natexlab}[1]{#1}
\providecommand{\url}[1]{\texttt{#1}}
\providecommand{\urlprefix}{}

\bibitem[{{GRAYSON} et~al.(2004)A. C. R. {GRAYSON} and R. S. {SHAWGO} and A. M.
  {JOHNSON} and N. T. {FLYNN} and {YAWEN LI} and M. J. {CIMA} and R.
  {LANGER}}]{10.1109/JPROC.2003.820534}
{GRAYSON} ACR, {SHAWGO} RS, {JOHNSON} AM, {FLYNN} NT, {YAWEN LI}, {CIMA} MJ,
  et~al.
\newblock A BioMEMS review: MEMS technology for physiologically integrated
  devices.
\newblock Proceedings of the IEEE 2004;92(1):6--21.

\bibitem[{Whitesides(2006)Whitesides, George M.}]{10.1038/nature05058}
Whitesides GM.
\newblock {The origins and the future of microfluidics}.
\newblock Nature 2006;442(7101):368--373.
\newblock \urlprefix\url{https://doi.org/10.1038/nature05058}.

\bibitem[{Dittrich et~al.(2006)Dittrich, Petra S. and Tachikawa, Kaoru and
  Manz, Andreas}]{10.1021/ac0605602}
Dittrich PS, Tachikawa K, Manz A.
\newblock Micro Total Analysis Systems. Latest Advancements and Trends.
\newblock Analytical Chemistry 2006;78(12):3887--3908.
\newblock \urlprefix\url{https://doi.org/10.1021/ac0605602}, pMID: 16771530.

\bibitem[{Wu et~al.(2010)Wu,Banqiu and Kumar,Ajay and
  Pamarthy,Sharma}]{10.1063/1.3474652}
Wu B, Kumar A, Pamarthy S.
\newblock High aspect ratio silicon etch: A review.
\newblock Journal of Applied Physics 2010;108(5):051101.
\newblock \urlprefix\url{https://doi.org/10.1063/1.3474652}.

\bibitem[{{Franz Laermer, Andrea Schilp}(granted on 26 March 1996){Franz
  Laermer, Andrea Schilp}, Robert Bosch Gbmh.}]{BoschPatent}
{Franz Laermer, Andrea Schilp} RBG, {Method of anisotropically etching
  silicon}; granted on 26 March 1996.
\newblock \urlprefix\url{https://patents.google.com/patent/US5501893A/en}.

\bibitem[{Laermer and Urban(2019)Laermer, Franz and Urban,
  Andrea}]{10.1002/ppap.201800207}
Laermer F, Urban A.
\newblock {MEMS at Bosch – Si plasma etch success story, history,
  applications, and products}.
\newblock Plasma Processes and Polymers 2019 sep;16(9):1800207.

\bibitem[{Laermer et~al.(2020)Laermer, Franz and Franssila, Sami and Sainiemi,
  Lauri and Kolari, Kai}]{10.1016/B978-0-12-817786-0.00016-5}
Laermer F, Franssila S, Sainiemi L, Kolari K.
\newblock {Deep reactive ion etching}.
\newblock In: Handbook of Silicon Based MEMS Materials and Technologies
  Elsevier; 2020.p. 417--446.

\bibitem[{Chen et~al.(2007)Chen, S. C. and Lin, Y. C. and Wu, J. C. and Horng,
  L. and Cheng, C. H.}]{10.1007/s00542-006-0211-2}
Chen SC, Lin YC, Wu JC, Horng L, Cheng CH.
\newblock {Parameter optimization for an ICP deep silicon etching system}.
\newblock Microsystem Technologies 2007 mar;13(5-6):465--474.
\newblock
  \urlprefix\url{https://link.springer.com/article/10.1007/s00542-006-0211-2}.

\bibitem[{Chung(2004)Chung, Chen-Kuei}]{10.1088/0960-1317/14/4/029}
Chung CK.
\newblock {Geometrical pattern effect on silicon deep etching by an inductively
  coupled plasma system}.
\newblock Journal of Micromechanics and Microengineering 2004
  apr;14(4):656--662.
\newblock
  \urlprefix\url{https://iopscience.iop.org/article/10.1088/0960-1317/14/4/029}.

\bibitem[{Lai et~al.(2006)S. L. Lai and D. Johnson and R.
  Westerman}]{10.1116/1.2172944}
Lai SL, Johnson D, Westerman R.
\newblock Aspect ratio dependent etching lag reduction in deep silicon etch
  processes.
\newblock Journal of Vacuum Science {\&} Technology A: Vacuum, Surfaces, and
  Films 2006 Jul;24(4):1283--1288.
\newblock \urlprefix\url{https://doi.org/10.1116/1.2172944}.

\bibitem[{Fu et~al.(2001)Fu, L. and Miao, J.M and Li, X.X and Lin,
  R.M}]{10.1016/S0169-4332(01)00197-0}
Fu L, Miao JM, Li XX, Lin RM.
\newblock {Study of deep silicon etching for micro-gyroscope fabrication}.
\newblock Applied Surface Science 2001 jun;177(1-2):78--84.
\newblock
  \urlprefix\url{https://linkinghub.elsevier.com/retrieve/pii/S0169433201001970}.

\bibitem[{Laermer and Urban(2003)Laermer, F. and Urban,
  A.}]{10.1016/S0167-9317(03)00089-3}
Laermer F, Urban A.
\newblock {Challenges, developments and applications of silicon deep reactive
  ion etching}.
\newblock Microelectronic Engineering 2003 jun;67-68:349--355.
\newblock
  \urlprefix\url{https://linkinghub.elsevier.com/retrieve/pii/S0167931703000893}.

\bibitem[{Abdolvand and Ayazi(2008)Abdolvand, Reza and Ayazi,
  Farrokh}]{10.1016/j.sna.2007.12.026}
Abdolvand R, Ayazi F.
\newblock {An advanced reactive ion etching process for very high aspect-ratio
  sub-micron wide trenches in silicon}.
\newblock Sensors and Actuators A: Physical 2008 may;144(1):109--116.
\newblock
  \urlprefix\url{https://linkinghub.elsevier.com/retrieve/pii/S0924424708000046}.

\bibitem[{Parasuraman et~al.(2014)Parasuraman, Jayalakshmi and Summanwar, Anand
  and Marty, Fr{\'{e}}d{\'{e}}ric and Basset, Philippe and Angelescu, Dan E.
  and Bourouina, Tarik}]{10.1016/j.mee.2013.06.010}
Parasuraman J, Summanwar A, Marty F, Basset P, Angelescu DE, Bourouina T.
\newblock {Deep reactive ion etching of sub-micrometer trenches with ultra high
  aspect ratio}.
\newblock Microelectronic Engineering 2014;113:35--39.
\newblock \urlprefix\url{http://dx.doi.org/10.1016/j.mee.2013.06.010}.

\bibitem[{Morton et~al.(2008)Morton, Keith J. and Nieberg, Gregory and Bai,
  Shufeng and Chou, Stephen Y.}]{10.1088/0957-4484/19/34/345301}
Morton KJ, Nieberg G, Bai S, Chou SY.
\newblock {Wafer-scale patterning of sub-40 nm diameter and high aspect ratio
  ({\textgreater}50:1) silicon pillar arrays by nanoimprint and etching}.
\newblock Nanotechnology 2008 aug;19(34):345301.
\newblock
  \urlprefix\url{https://iopscience.iop.org/article/10.1088/0957-4484/19/34/345301}.

\bibitem[{Owen et~al.(2012)Owen, K. J. and VanDerElzen, B. and Peterson, R. L.
  and Najafi, K.}]{10.1109/MEMSYS.2012.6170138}
Owen KJ, VanDerElzen B, Peterson RL, Najafi K.
\newblock {High aspect ratio deep silicon etching}.
\newblock In: 2012 IEEE 25th International Conference on Micro Electro
  Mechanical Systems (MEMS) No. February, IEEE; 2012. p. 251--254.
\newblock \urlprefix\url{http://ieeexplore.ieee.org/document/6170138/}.

\bibitem[{Blauw et~al.(2001)Blauw, M. A. and Zijlstra, T. and van der Drift,
  E.}]{10.1116/1.1415511}
Blauw MA, Zijlstra T, van~der Drift E.
\newblock {Balancing the etching and passivation in time-multiplexed deep dry
  etching of silicon}.
\newblock Journal of Vacuum Science {\&} Technology B: Microelectronics and
  Nanometer Structures 2001;19(6):2930.

\end{thebibliography}
\section*{Table of Contents}
\begin{figure}[h!]
\centering
    \includegraphics{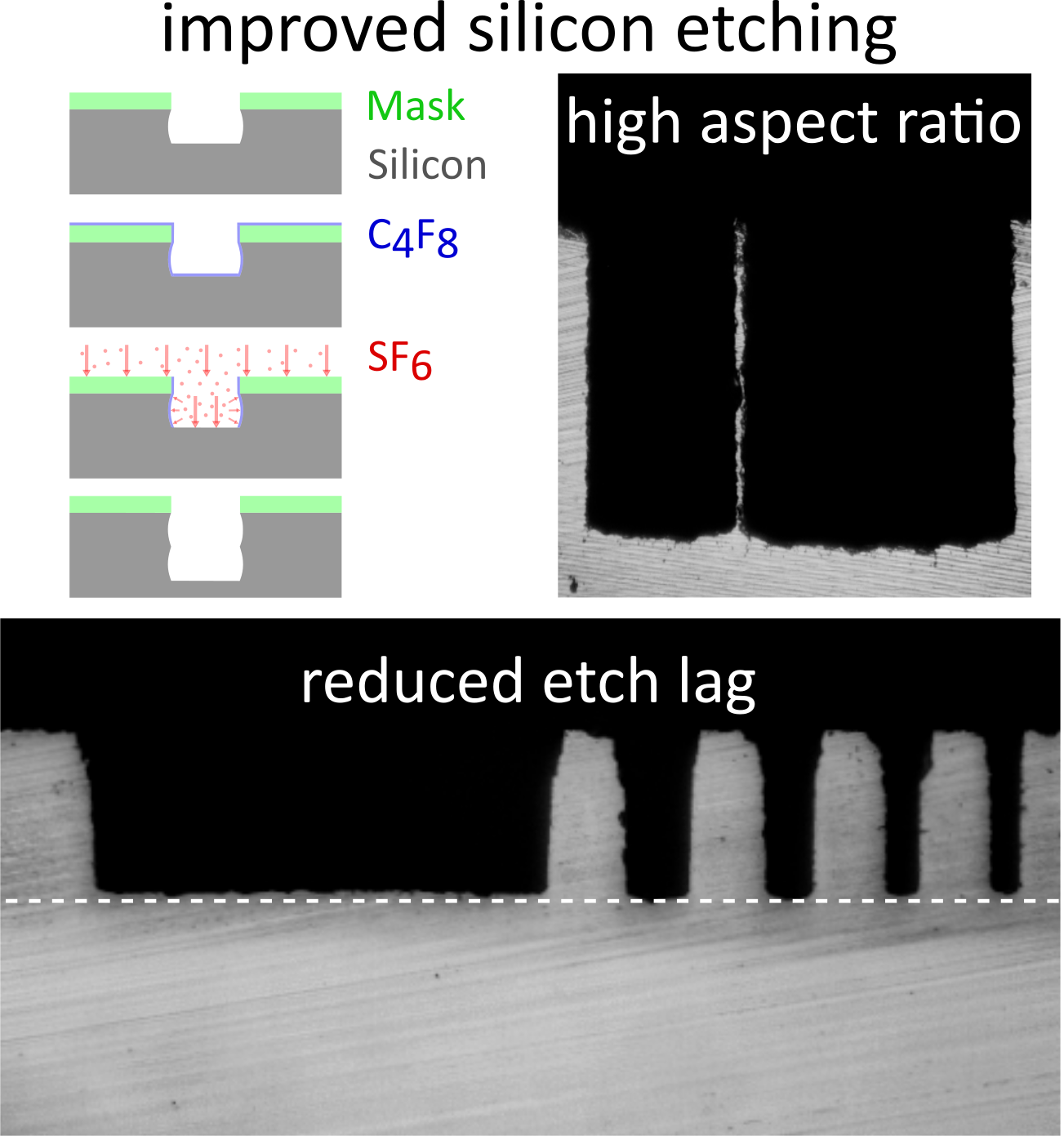}
\end{figure}
   
Reactive ion etching with the Bosch process is one of the key procedures used to manufacture micron-sized structures for MEMS and microfluidic applications. We introduce an optimized Bosch process that reduces the etch lag to below $\SI{1.5}{\percent}$. Furthermore, we improved a three-step Bosch process, allowing the fabrication of $\SI{6}{\micro\meter}$ thin structures at depths up to $\SI{200}{\micro\meter}$ while maintaining their stability.



\end{document}